\documentclass[preprint,aps,pre]{revtex4-1}  
\usepackage{graphicx}
\usepackage{amsmath,amssymb}
\usepackage{bm}
\usepackage{tcolorbox}
\usepackage{hyperref}

\def\bea{\begin{eqnarray}}
\def\eea{\end{eqnarray}}

\def\nn{\nonumber}

\def\f{\frac}

\begin{document}

\title{ A critique of the Covid-19 analysis for India by Singh and Adhikari }
\author{Abhishek Dhar}
\affiliation{International Center for Theoretical Sciences, Tata Institute of Fundamental Research, Bangalore-560089, India}

\begin{abstract}
In a recent paper \cite{adhikari2020}, Singh and Adhikari present results of an analysis of a mathematical model of epidemiology based on which they argue that a $49$ day lockdown is required in India for containing the pandemic in India. We assert that as a model study with the stated assumptions, the analysis presented in the paper is perfectly valid, however, any serious comparison with real data and attempts at prediction from the model are  highly problematic. The main point of the present paper is to convincingly establish this assertion while providing a warning that the results and analysis of such mathematical models should  not be taken at face value and need to be used with great caution by policy makers.
\end{abstract}

\maketitle

\section{Introduction}

A large number of recent papers have analyzed mathematical models, of varying degrees of complexity, in an attempt to understand and sometimes make predictions about the growth and spread of Covid-19 across the world. Such models often come with serious limitations arising from several  facts such as: (i) compartmentalization of the actual degrees of freedom to make analysis simpler, (ii) too many parameters whose values are known with high degrees of uncertainties, (ii) inherent nonlinearities in the governing equations which make long time predictions difficult. 

Nevertheless, there have been  papers which make (or appear to make) definitive predictions and many of these end up generating a lot of media hype and get wide attention in the social media. One real risk is that this is actually noticed and used by policy makers. In the present note, we analyze the paper by Singh and Adhikari, which we believe  is in this class. In particular, the paper appears to make the claim that a $49$ day lockdown is required to effectively control the growth of the Covid-19 pandemic in India.  In this note, we explain why this claim should not be taken seriously. 

At the outset we make it clear that the present comment does not question the technical validity of the results in \cite{adhikari2020}.  However the present comment points out that there are many issues on details such as that of interpretation of variables while making comparison with real data and choice of model parameters --- these make any attempts at predictions meaningless.

In Sec.~(II) we first analyze a simpler version of the model studied in \cite{adhikari2020} and show that for a certain choice of model parameters, this already reproduces the main features seen in the more complicated model. However we show that, on making small changes of the parameters to somewhat more realistic values, the predictions change significantly. In  Sec.~(III) we point out a large number of other problems related to the study in \cite{adhikari2020} which  imply that  numerical predictions emerging from such studies are completely unreliable.

\section{An analysis of a simplified version of the SIR-type model of Singh and Adhikari}

The basic model considered by the authors is one which compartmentalizes  the population according to age. At any time each compartment has a number of susceptible ($S$) ,  
 infected ($I$) and recovered ($R$) individuals and these evolve with time. The division into age groups is to account for the fact that the levels of contact between  different groups (and within groups) vary and so they have different probabilities of passing infections. To discuss the main aspects of the model, it is sufficient to ignore the compartmentalization into age groups and we  present the simpler  version here.

This more basic  model is one which divides the population  of size $N$ 
into individuals who are susceptible ($S$) , asymptomatic infectious ($I^a$), symptomatic infectious ($I^s$) or are recovered ($R$), with the  constraint $N=S+I^a+I^s+R$. The number $R$ also includes people who have died.

{\bf Dynamics}: Let $I=I^a+I^s$ be the total number of infected individuals. For a well mixed population, the probability that an $S$ individual meets an $I$  individual  is  $\propto (I/N)$. We define the infection rate $\beta$,  which controls the rate of spread and represents the probability of transmitting disease between a susceptible and an infectious individual.  We assume that infected people typically either recover or die after $T_R$ days, so $I \to R$ happens at a rate $\gamma = 1/T_R$. We also assume that a fraction $\alpha$ of the infected population is asymptomatic. However, as in the analysis of \cite{adhikari2020}, we will set $\alpha=0$ which effectively means that we ignore asymptomatic carrier populations (if any).

We then have  the following dynamics for the system
\begin{align}
\f{dS}{dt}&=-\f{\beta u (t) I}{N} S  \\
 \f{dI^a}{dt}&=\alpha \f{\beta u(t) I}{N} S  -\gamma I^a\\
\f{dI^s}{dt}&=(1-\alpha) \f{\beta u(t) I}{N} S  -\gamma I^s\\
\f{dR}{dt}&=\gamma I, \label{sir}
{\rm where~~} I =I^a +I^s.
\end{align} 
The time dependent function $u(t)$ is the ``lockdown'' function that incorporates the effect of a lockdown on the rate of spreading of the infection. A reasonable form is one where $u(t)$ has the constant value $(=1)$  before the beginning of the lockdown,  at time $t_{on}$, and then  it changes to $ 0< u_l <1$ over a characteristic time scale  $\sim t_w$. Thus we can take a form
\begin{align}
u(t)&=1~~~t<t_{\rm on}, \nn \\ 
&=u_l+(1-u_l)e^{-(t-t_{\rm on})/t_w},~~~t>t_{\rm on}. 
\end{align}
 The number $u_l$ indicates the lowering of social contacts as a result of a lockdown. This was set to the value $0$ in \cite{adhikari2020} but we note that realistically it is expected that social contacts cannot be brought down to zero (for example, people still have to get groceries). 

An important parameter is the so-called  reproductive number  $R_0=\beta/\gamma$ which is the average number of new people infected by one infected person. Typical values reported in the literature for  Covid-19 are in the range $R_0=2-7$ \cite{reviewR0}.  Note that  while $R_0$ is like the ``free'' reproductive number, $u(t) R_0$ is  an effective reproductive number incorporating the fact that the number of contacts any individual makes has been brought down as a result of the lock down.

The other parameters in the model are $\gamma, \alpha$ and the size of the susceptible population itself. Apart from these parameters, a solution of the above equations requires the initial data for $S(0), I^a(0), I^s(0)$ and $R(0)$.   

An immediate property of the model is that it leads to an exponential growth of the infected population at intermediate times. We can understand this easily by a linear analysis around the initial state with $S=N, I^a=I^s=R=0$. This shows that the infected numbers grow exponentially as $\sim e^{\lambda t}$ with 
$\lambda=\beta-\gamma$.  Thus we get a growth only if $R_0=\beta/\gamma >1$. 

We now show results from the numerical solution of these equations with a choice of parameters ({\bf Set I} below) consistent with \cite{adhikari2020} and make a comparison of  our results with theirs.  We find that the predictions of the simplified model are more or less the same as those obtained from the more complicated model. However, on changing parameters to somewhat more realistic values [{\bf Sets (II,III)}], the predictions change drastically. 
\begin{figure}{*}
\center
\includegraphics[scale=0.6]{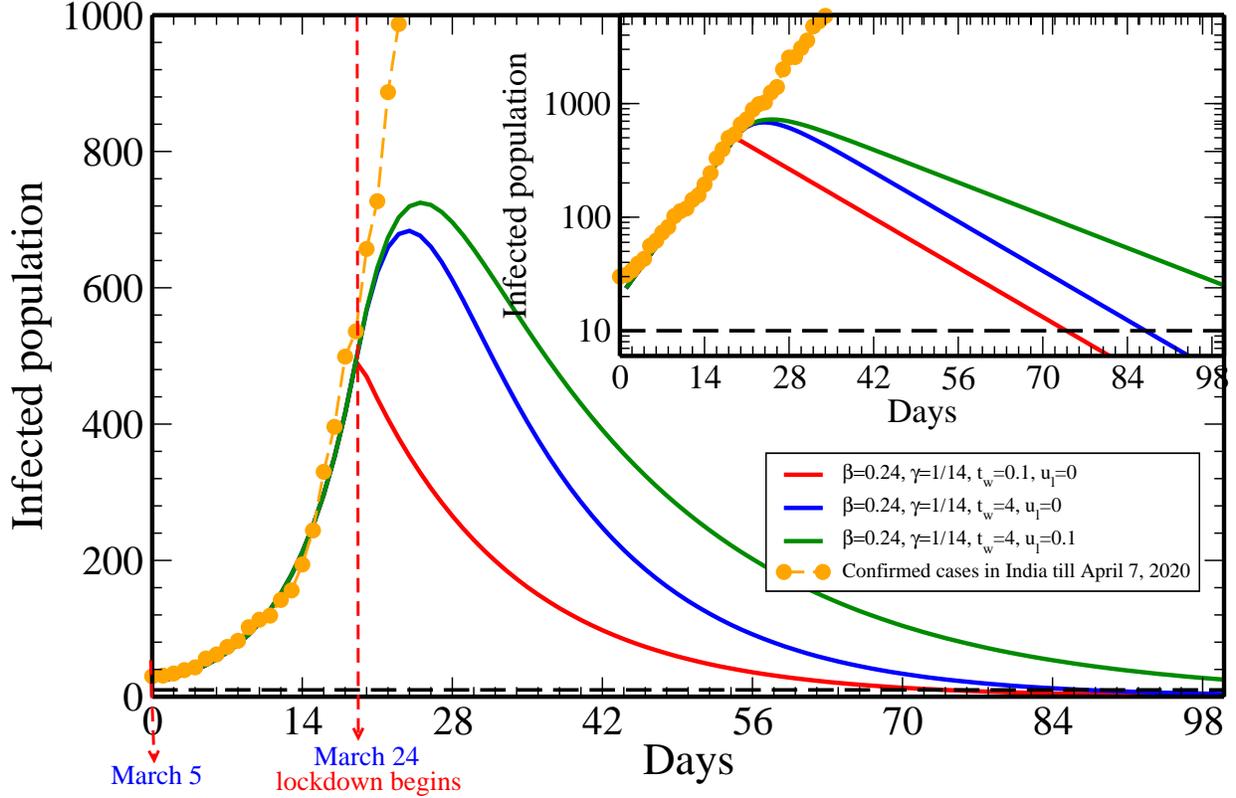}
\caption{{\bf Forecast of the COVID-19 epidemic in India with the simplified SIR model}: The evolution of the SIR model in Eqs.~(1-4) starting with a seed infected population of size $I(0)=20$ and a total population $N=10^9$. Lockdown is imposed after $t_{\rm on}= 19$ days. The other parameter sets are given in the figure legend. The horizontal black dashed line indicates an infected population of size $I=10$ and we see that this number is reached approximately after $54$ days from the start of lockdown (for Set I), after $68$ days for Set II, while for Set III this takes about  $101$ days. The evolution for parameter set (I) is very close to the  results presented in \cite{adhikari2020}. Note also that the evolution for the three parameter sets are identical before the lockdown. \emph{Notably, the real data (for the period March 6 - April 7, 2020) is following a completely different trajectory (orange points)}. The inset shows the same  data plotted on a semi-log scale.}
\label{sirfig}
\end{figure}

{\bf Parameter set (I)}: We choose the parameter values $\gamma=1/14$ per day ($14$ days recovery time) and $\beta=0.24$ per day. This gives $R_0=3.36$ and the  observed  exponential growth rate of around $\lambda \approx 0.17$ (doubling time $\approx 4$ days). We also fix $\alpha=0$, $t_w=0.1$ and  $u_l=0$. With this choice of parameters, let us now see how the infected population evolves as a function of time for the choice of initial conditions $I^s=20$, $I^a=R=0$ $N=10^9$. [{\bf Notes}: (a) the values of $\beta$ and $\gamma$ are different from \cite{adhikari2020} but they are consistent with known estimates of $R_0$ and the observed exponential growth rate in the Indian data for confirmed cases, (b) the values of $\alpha, t_w$ and $u_l$ are as those used by \cite{adhikari2020} (they take $t_w<1$) and are clearly unrealistic, (c) the initial conditions are chosen so as to fit the initial growth seen in the real data after the confirmed number of cases is $\sim 20$.] 

{\bf Parameter set (II)}: We keep  $\gamma=1/14, \beta=0.24, \alpha=0$ and $u_l=0$ but change the time, for the lockdown to be effective, to a more realistic value of $t_w=4.0$ days.

{\bf Parameter set (III)}: We keep  $\gamma=1/14, \beta=0.24, \alpha=0$  but change both the time, for the lockdown to be effective, to $t_w=4.0$ days and take $u_l=0.1$ (since, as discussed above, contacts are never reduced to $0$).

In Fig.~(1) we show the results for these three parameter sets and a comparison with real data for the number of confirmed Covid-19 cases in India. 
\noindent [{\bf Note}: For comparison with real data, we follow \cite{adhikari2020} and assume that $I^s(t)$ is equal to the number of reported confirmed cases.
In the comments in the next section, we explain however that  this is \emph{not} a reasonable assumption.]

In Fig.~(1) we  observe that the  results from parameter set (I) are already in close agreement to those  presented in Fig.~(4d) of \cite{adhikari2020}.  The estimated lockdown period for the number of infected to reduce to a value less than $10$ is $54$ (which compares remarkably well with the number $49$ in \cite{adhikari2020}, the small difference is expected in this type of analysis).  
For the  three parameter sets we find that the time required for the infected population to reduce to the value $I=10$ are respectively 
\begin{align}
Set~(I) \to ~54 ~{\rm days}, \nn \\
Set~(II) \to ~68~{\rm  days}, \nn \\
Set~(III) \to 101~{\rm  days}. 
\label{lockdown}
\end{align}
We thus conclude that small changes in parameters which make the model more realistic change the predictions significantly. More importantly, as is clear from Fig.~(1) and the inset, the predictions of the model are nowhere following the actual trajectory of the pandemic. This is of course also true for the predictions of the more sophisticated model studied in \cite{adhikari2020} --- \emph{the true data after the lockdown has no resemblance to their predicted evolution}.  

{\bf Warning}: The improved estimates quoted  in Eq.~\eqref{lockdown} should not be used as suggested periods of lockdown. They are given here simply to point out the limitations of the present model. 

We end this section by emphasizing that while the analysis presented  here ignores details such as the contact population's age break up, population distribution and connectivity matrices, we already recover the main qualitative (and to a large extent, even quantitative) aspects of the results in \cite{adhikari2020}. It is expected that the main message on sensitivity of model predictions to parameter choices would not change by increasing the complexity of the model.

\section{A list of  problems with the paper}

In the previous section, through the analysis of a simpler model,  we highlighted some of the reasons why models such as those studied in \cite{adhikari2020} cannot be taken seriously for their predictive powers. However there are many more serious problems in the analysis and claims of the paper. Some of these are:
\begin{enumerate}

\item The modeling ignores important factors such as latency periods, asymptomatic and presymptomatic transmission. Presymptomatic transmission could  happen over the order of $5-10$ days and is a period during which individuals show no symptoms (and are undetected) but are infecting others. 

\item The modeling ignores the fact that the number of people who have been tested are relatively very small in India and the actual number of infected people could be much higher.

\item A related point  is the following ---  in their study, the number of infected individuals $I$ is compared with the total reported number of confirmed cases. However one expects that the reported confirmed cases (say  $C$) are those that habve  already been detected and therfore  it does not make sense  that they should be counted amongst the ones who are spreading the infection. It seems reasonable to us that $I >> C$. So the comparison with real data ($C$) through the  identification $I=C$ is  flawed. In fact it seems to us that it makes more sense to instead compare $R$ with the real data.

\item The choice of initial conditions in solving these equations is subtle and it is unclear how the  authors fix these. Small changes in initial conditions can lead to drastically different predictions. 

\item As we have demonstrated, small changes in parameter values would lead to drastically different predictions.

\item The choice of the lockdown parameters $t_w<1$ and $u_l=0$ are completely unrealistic.  We expect there should be a time lag for the lockdown to take effect and even under lockdown, social contacts are not reduced to zero.

\item  The optimal lockdown period is computed based on the arbitrary criterion that the infected number drops to the value $10$. Why not $100$  which would give a completely different prediction for the required lockdown period ? We do not see why this completely arbitrary value should be used as a criterion.

\item The actual trajectory of infections (or rather number of confirmed cases)  after the lockdown started in India has no resemblance whatsoever with what is predicted by the authors, even for short times. In fact, a data analysis of the effect of lockdowns in several countries on the confirmed number of cases, presented in \cite{website}, shows that the evolution differs \emph{qualitatively} in all the countries from the picture presented in \cite{adhikari2020}.  \emph{In our opinion, this basically proves that the model has no predictive value and should not be used as  guiding material for  policy makers}.
\end{enumerate}

\section{Conclusions}
We have established in this brief note that the  studies of mathematical models of disease spreading,  such as the one in \cite{adhikari2020},  typically have { very limited  ability to make numerical predictions}.  Such models, if analyzed properly,  could play an important  role in our qualitative understanding  of disease evolution and, with some effort, perhaps in making short term predictions. Unfortunately, completely absurd claims sometimes attract a lot of public attention and  run the risk of leading  to wrong  policy decisions. 

While there are many such examples in the recent literature, we have focused here on analyzing one specific model study \cite{adhikari2020}. This paper is technically correct in that they state their assumptions clearly and the stated results are obtained under these assumptions. However, we have  pointed out a number of serious problems in their analysis [listed in Sec.~(III) of the present paper] which makes any numerical predictions from the model for the growth of the pandemic in India completely meaningless. This becomes obvious when one compares the
real growth curve, since the date of the lockdown, with the predicted one [see Fig.~(1)]. The model fails, even at a qualitative level, right from  the  time of the lockdown.

\begin{acknowledgements}
I thank Suvrat Raju for first flagging possible problems with the paper. 
I thank Arghya Das, Srashti Goyal, Jitendra Kethepalli and Kanaya Malakar for extensive discussions on this topic.  I am grateful to Ranjini Bandyopadhyay, Debasish Chaudhuri, Chandan Dasgupta, Anupam Kundu, Satya Majumdar,  Samriddhi Sankar Ray and Sanjib Sabhapandit  for a careful reading of the manuscript and pointing out errors. 
I thank  Rajesh Gopakumar, Sandeep Juneja,  Sandeep Krishna, Vijay Kumar Krishnamurthy, Gautam Menon, Pramod Pullarkat,  Sriram Shastry, Rajesh Sundaresan, Sumati Surya, Mukund Thattai and Spenta Wadia  for their comments on the manuscript.
\end{acknowledgements}

\end{document}